\documentclass[sigconf]{acmart}
\usepackage[utf8]{inputenc}

\usepackage{rotating}
\usepackage{tabu}

\usepackage{subcaption}

\copyrightyear{2018} 
\acmYear{2018} 
\setcopyright{acmlicensed}
\acmConference[WebSci '18]{10th ACM Conference on Web Science}{May 27--30, 2018}{Amsterdam, Netherlands}
\acmBooktitle{WebSci '18: 10th ACM Conference on Web Science, May 27--30, 2018, Amsterdam, Netherlands}
\acmPrice{15.00}
\acmDOI{10.1145/3201064.3201101}
\acmISBN{978-1-4503-5563-6/18/05}

\acmConference[WebSci'18]{ACM Conference on Web Science}{May 2018}{Amsterdam}
\acmYear{2018}
\copyrightyear{2018}

\acmArticle{4}
\acmPrice{15.00}

\begin{document}
\title{Can We Count on Social Media Metrics?%Non-Tra\-ditional Re\-search Me\-trics?\\
\\
First In\-sights into the Active Scholar\-ly Use of Social Media }

\author{Maryam Mehrazar
}
\affiliation{%
  \institution{\mbox{ZBW Leibniz Information Centre}}
  \streetaddress{ D"usternbrooker Weg 120 }
  \city{for Economics, Kiel}
  \state{Germany}
  \postcode{24105}
}
\email{m.mehrazar@zbw.eu}

\author{Christoph Carl Kling}
\affiliation{%
  \institution{www.c-kling.de}
  \city{Cologne}
  \state{Germany}
  \postcode{50825}
}
\email{datascience@c-kling.de}

\author{Steffen Lemke
}
\affiliation{%
  \institution{\mbox{ZBW Leibniz Information Centre}}
  \streetaddress{ D"usternbrooker Weg 120 }
  \city{for Economics, Kiel}
  \state{Germany}
  \postcode{24105}
}
\email{s.lemke@zbw.eu}

\author{Athanasios Mazara\-kis
}
\affiliation{%
  \institution{Kiel University}
  \streetaddress{}
  \city{\hspace{0cm} Kiel}
  \state{Germany}
  \postcode{24118}
}
\email{a.mazarakis@zbw.eu}

\author{Isabella Peters}
\affiliation{%
  \institution{\mbox{ZBW Leibniz Information Centre}}
  \streetaddress{ D"usternbrooker Weg 120 }
  \city{for Economics, Kiel}
  \state{Germany}
  \postcode{24105}
}
\email{           i.peters@zbw.eu}

%\author{Steffen Lemke}
%\affiliation{%
%  \institution{ZBW -- Leibniz Information Centre for Economics}
%  \streetaddress{ D"usternbrooker Weg 120 }
%  \city{Kiel}
%  \state{Germany}
%  \postcode{24105}
%}
%\email{s.lemke@zbw.eu}

%\author{Isabella Peters}
%\affiliation{%
%  \institution{ZBW -- Leibniz Information Centre for Economics}
%  \streetaddress{ D"usternbrooker Weg 120 }
%  \city{Kiel}
%  \state{Germany}
%  \postcode{24105}
%}
%\email{i.peters@zbw.eu}

\renewcommand{\shortauthors}{M. Mehrazar et al.}

%Scholarly metrics which measure research impact are important for improving academic search engines and for research evaluation.
%\textit{Social media metrics} or \textit{altmetrics}, measure the impact of scientific work based on social media activity, e.g.\ allowing the real-time evaluation of new publications. 
%

\begin{abstract}
%Measuring the impact of research using scholarly metrics is important for improving the performance of academic search engines as well as for hiring decisions and project evaluation.
%Measuring the impact of research using scholarly metrics is an important part of today's scientific process. 
%\textit{Social media metrics}, also known as \textit{altmetrics}, measure the impact of scientific work based on social media activity. 
%These non-traditional metrics are complementary to traditional, citation-based metrics. %and are gaining popularity recently.
Measuring research impact is important for ranking publications in academic search engines and for research evaluation.
\textit{Social media metrics} or \textit{altmetrics} measure the impact of scientific work based on social media activity.
Altmetrics are complementary to traditional, citation-based metrics, e.g.\ allowing the assessment of new publications for which citations are not yet available. %and are gaining popularity recently.

Despite the increasing importance of altmetrics, their characteristics are not well understood:
Until now it has not been researched what kind of researchers are actively using which social media services and why -- important questions for scientific impact prediction. %better understanding the scientific impact of scholarly work. 
%the weighting of activities for measuring scholarly impact.
%weighting activities when calculating scores.(*better rephrase: for finding the most important activities for the impact assessment)
%behaviour of social media user and whether we mistake affordances for intentional interaction with scholarly products,
%lack of understanding of user activities (e.g. motivations to retweet, like, share) and user groups in academic contexts
Based on a survey among 3,430 scientists, we 
%shed light on the demographics of researchers actively using social media, and 
uncover previously unknown and significant differences between %the demographics and motivations of active scholarly users of 
social media services: 
We identify services which attract young and experienced researchers, respectively, and detect differences in usage motivations. 
Our findings have direct implications for the future design of altmetrics for scientific impact prediction.%improved social media metrics.
\end{abstract}


%\keywords{social media, digital scholarship, altmetrics, motivations}

\maketitle

\section{Introduction}
%In the following we will present some preliminary results on the specifities of altmetrics provided by scholars that use social media.

The use of the web is an integral part of scientific work. 
On social media, researchers discover new research, discuss research ideas with fellows and disseminate research results to the public
and to the scientific community~\cite{priem2011altmetrics,hadgu2014identifying,conf/websci/Jordan15}. 
Additionally, academic search engines support scientists in finding scholarly literature.

In order to improve their performance, academic search engines employ \textit{scholarly metrics}: citation-based measures for the scientific impact of authors and scientific works~\cite{sugimoto2017scholarly}.
In fact, scholarly metrics are also important for other applications such as hiring decisions and project and application evaluation~\cite{niso,beel2009google}.

One drawback of traditional, citation-based metrics is that citations are not available for new publications -- the first citation of a paper may take years.
Additionally, scholarly metrics do not cover %the scientific process on the web, which is becoming increasingly important.
the scientific impact on the web. %, which is still gaining importance. 
Therefore, \textit{social media metrics} or \textit{altmetrics} were introduced as a complement to traditional metrics: By analysing usage patterns on social media, altmetrics evaluate the quality of scholarly products through their impact on the web~\cite{priem2011altmetrics}.
Altmetrics which predict the \textit{scientific} impact of scholarly work~\cite{zoller2015publication} will likely play a central role in many future applications such as scientific literature retrieval.

To the best of our knowledge, current altmetrics data providers such as \textit{altmetric.com}~\cite{priem2011altmetrics,DBLP:journals/corr/Priem15} or \textit{PlumX}~\cite{champieux2015plumx} use sums or simplistic weightings for aggregating altmetrics %activity counts related to scholarly works 
(e.g.\ the number of mentions) from different social media services. For instance, view counts are aggregated across services using weighted sums. 
It has not yet been investigated whether this practice reflects the complexity of actions on social media. 
In order to improve altmetrics for scientific impact prediction, it is essential to understand the demographics and motives of scholarly social media users. 
%simplistic weightings
If social media services differ significantly in the demographics or motives of their users, the mechanisms of altmetrics would have to be improved: One example could be a service-specific correction for the share of postdocs, who are known to have a high productivity~\cite{carayol2006individual} and thus create more citations, which are to be predicted.

This paper analyses the results of a survey among 3,430 scientists, providing first insights into the scholarly use of social media by detecting and describing
\textbf{(i)~demographic differences of active scholarly users of social media} and
\textbf{(ii)~variations in the motivation for scholarly use of social media} between services.

It is well-known that a small share of active users in social media contributes the majority of observed activities, the so-called ``90:9:1 rule''~\cite{nielson2006alertbox}. 
%And it is precisely these activities that altmetrics measure. 
As a result, active users are responsible for most of the activities measured by altmetrics.
%These active users dominate altmetrics that develop around scholarly works.
%These active users dominate in creating altmetrics for scholarly works.
%And it is precisely these activities that are measured by altmetrics.
Unlike previous analyses of the scholarly use of social media~\cite{conf/websci/Jordan15,zoller2015publication}, \textbf{we therefore only consider \textit{active} users} 
%for our analyses
who use social media at least weekly for scientific purposes.

\begin{figure*}[!t]
    \centering
    \includegraphics[width=\textwidth]{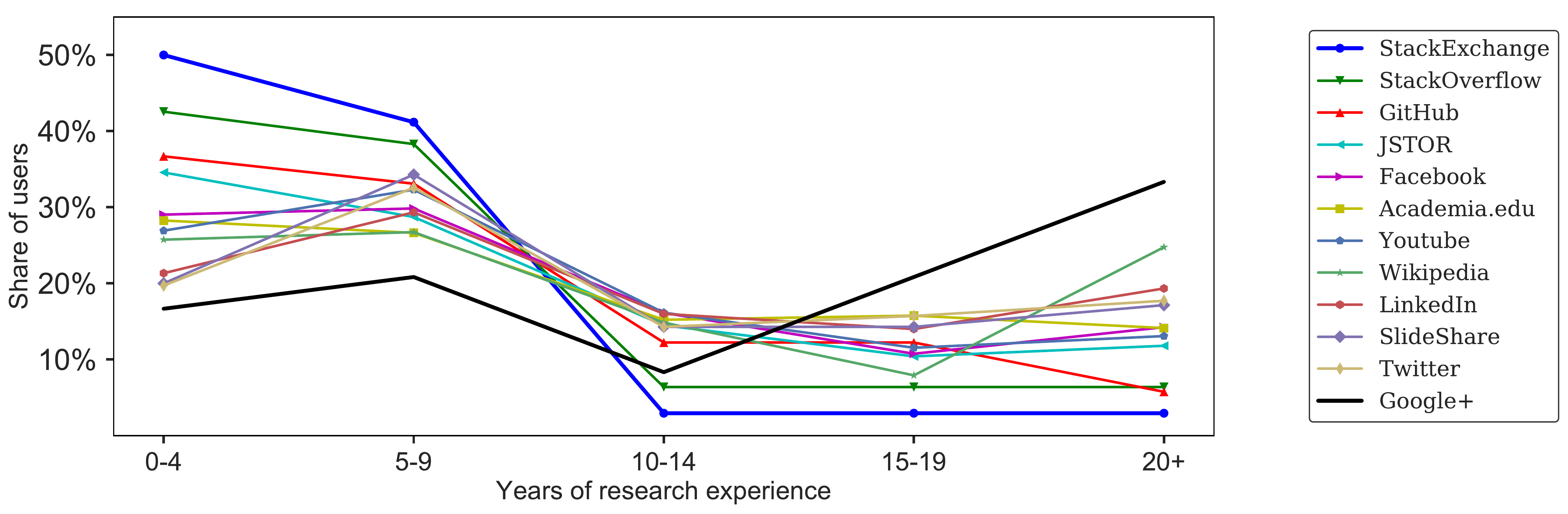}
    \caption{\textbf{Academic experience of active users per service.} \textnormal{%The experience of users varies greatly. 
    Programming-related services (StackExchange, StackOverflow, GitHub) are used by young scientists, while networking services (Google+, Twitter, LinkedIn), SlideShare and Wikipedia have the highest percentage of experienced scientists. The legend shows services ordered by their share of young scientists (0-4 years of experience). The services with the highest share of young researchers (StackExchange) and the highest share of experienced academics (Google+) are highlighted. Social media services show substantial differences in the experience level of active scholarly users.}}
    \label{fig:exp_per_service}
\end{figure*}

%%%%%%%%%%%%%%%%%%%%%%%%%%%%%%%%%%%%%%%%%%%%%%%%%%%%%%%%%%%%%%%%%%%%%%%%%%%%%%%%%%%%%%%%%%%%%%%%%%%%%%
\section{Related Work}

Social media have become increasingly popular for scholarly communication~\cite{ponte2011scholarly, tenopir2013social}. Several metrics based on scholarly social media activities have been shown to correlate with traditional, citation metrics~\cite{zoller2015publication, mohammadi2015reads}, though previous studies have pointed out that there is wide variability in the social media use of researchers. Differences have been observed in age, academic role, discipline and country, among others~\cite{hadgu2014identifying, procter2010adoption, nicholas2011social, bowman2015investigating, mikki2015digital}. 
%One study~\cite{nicholas2011social} found age to be a poor predictor of scholarly social media use, while other 
%Several studies found significant associations between age and scholarly use of social media~\cite{hadgu2014identifying, procter2010adoption, nicholas2011social, bowman2015investigating, mikki2015digital}. Academic roles also shape social media practices differently~\cite{procter2010adoption, dowling2017digital} and make it different for different types of platforms such as social bookmarking services such as Mendeley~\cite{haustein2014tweets}, micro-blogging services such as Twitter~\cite{bowman2015investigating} and blogs~\cite{shema2012research}. 

Using metrics based on social media activities comes with various challenges such as the assurance of data quality, the consideration of the heterogeneity of acts, users and motivations on social media, and the prevention of bias~\cite{haustein2016grand, bornmann2014altmetrics}. 
One key problem of scholarly social media data is the systematic bias towards scholars with certain demographic characteristics such as bias towards younger users~\cite{DBLP:journals/corr/Priem15} and towards users with a professional interest in research~\cite{neylon2009level}. 
Several studies state that the lack of accurate user statistics or sample descriptions for social media sites complicates the quantification of these biases~\cite{bornmann2014altmetrics,sugimoto2017scholarly}. 

Scholars use social media for various reasons. Van Noorden~\cite{van2014online} identified multiple categories of motivations for scholarly social media use:
%Ordered by the intensity of users' engagement, reasons for social media use include: 
\textit{contacting peers, posting content, sharing links to authored content, actively discussing research, commenting on research, following discussions, tracking metrics, discovering jobs, discovering peers, discovering recommended papers, offering a contact possibility} and \textit{curiosity}. %We found 6-7 of those in line with the motivations of active users...
%Disciplinary differences in Twitter scholarly communication \cite{holmberg2014disciplinary}
Jordan \cite{conf/websci/Jordan15} identified motivation categories by manually coding questions asked by researchers on Academia.edu.% However, both authors do not specifically analyse active users.
%These categorisation overlaps our findings. 

%what we know of scholarly use of social media. however they didn't didn't look at the active users.
It is well known that a small minority of active social media users is responsible for a large share of activities \cite{whittaker2003dynamics, kittur2007power}. Russo et al. ~\cite{russo2009great} give an overview of multiple studies 
%which confirm the effect 
on various social media sites and Kunegis~\cite{kunegis:konect-netsci} shows statistics for dozens of social networks, all confirming the effect.
To the best of our knowledge, there is no study on the demographics and motivations of \textit{active} scholarly users on social media.

\section{Data and Methods}
%As %a first step towards a deeper understanding of
%part of a larger research project on metrics, we conducted an exploratory   
%online survey\footnote{\url{https://tigereye.informatik.uni-kiel.de/limesurvey/index.php/847788}}
We analyse data from an exploratory 
online survey
%\footnote{\url{https://tigereye.informatik.uni-kiel.de/limesurvey/index.php/847788}}
\footnote{\url{https://github.com/marymm/-metrics/raw/master/questionnaire.pdf}}
%\footnote{\url{https://metrics-project.net/downloads/metrics_4am_poster.pdf}}
on the professional scholarly use of social media, 
which we conducted as part of a larger research project on metrics.
%\href{https://tigereye.informatik.uni-kiel.de/limesurvey/index.php/847788?lang=en}{online survey}
Among others, we asked participants for their research experience, their academic role, the social media services they use, and how often and why they are using social media services. %and how positive they rate the functionalities for the named services,
%and their motivation for using social media services.

\subsection{Survey data}

Our survey on social media usage was distributed via multiple channels: Authors who had at least one publication after 2015 with an email listed in the Web of Science, RePEc and multiple mailing lists related to Economics, Social Sciences or its subfields. As the survey was conducted as part of an interdisciplinary project involving partners from economics and social sciences, our main target group was economists and social scientists.

More than 3,430 international researchers participated in our survey from March to May 2017 with a response rate of about 6\%.
More than half of the participants -- 1,731 researchers -- use at least one functionality (e.g.\ like, share or post) of a social media service per week. We call these researchers \textit{active users}. 

%Our results are biased towards researchers who are experienced users and professors.
Most of the researchers are from the fields of economics (60\%) and social sciences (22\%). Researchers from 84 countries participated, the majority of them from Germany (51\%), followed by the US (10\%) and the UK (5\%). 
Ages of participants range from 19 to 89 (median age 38). %, and various academic roles were covered, including professors, postdocs and PhD students.
The distribution of academic roles is as follows: About 44\% of the participants are professors, followed by PhD students~/ research assistants (31\%) and postdocs (19\%).
This is in line with studies showing that professors together with PhD students have the highest share of profiles on academic social media%and represent the majority of users
~\cite{mikki2015digital, ortega2015academic}. 

%Apparently, experienced scientists and early career researchers are the most representative groups of scientists online and our survey is also clearly biased towards those. 
%We analyse the sample of the most active social media users from the survey and the sampled data

Though our sample is not representative, the high share of active users in our survey allows us to analyse differences between social media services.
If we find significant differences between services in our survey, we also expect to find differences in the parent population.

%Bias toward social scientists and economists, since they are our target group.
%Though our survey is clearly biased towards experienced scientists and early career researchers, 
%the data allows us to analyse differences between their behaviour and their motives for using social media: If we find significant differences in our survey, we also expect to find differences in a representative sample.

%not representative but
%first detailed analysis, though biased
%we look at differences between services -> if we find differences here, it is highly likely to find differences in the general population

\subsection{Experience differences and motivations}
In our survey, we asked participants to state their research experience since graduation using predefined ordinal categories (0-4 years etc.).
For detecting significant differences in the distribution of research experience, we look at all possible pairs between the twelve most-mentioned services and use pairwise $\chi^2$ tests on category counts of the answers of participants. %, at a significance level of $0.05$. We only considered pairs with strong effect sizes ($>0.25)$ and used Bonferroni correction.
We apply the Benjamini-Hochberg procedure with a false discovery rate of $0.05$ and only pairs with strong effect sizes ($>0.25)$ were considered.

Our survey contains a question on reasons for using social media. In order to detect latent motivations for using social media, we ran Latent Dirichlet Allocation 
(LDA)~\cite{Blei03ldaModel}, the most common topic model, on the free text answers. Topic models detect sets of semantically related words using the co-occurrence of words in documents.
%LDA finds hidden topics in the text by considering the co-occurrences of words in the answers. 
We chose to set the topic parameter to 10 topics (the lowest number yielding meaningful topics), and used sparse, symmetric document-topic and topic-word Dirichlet priors with $\alpha=\beta=0.1$.
Negative answers (e.g.\ ``none'') were manually deleted and stopwords (from NLTK~\cite{Loper02nltk:the}) were removed from the remaining answers, resulting in 997 answer texts.

\section{Results}
In this section, we look at differences between social media services in terms of demographics and motivations of active users. 

\begin{figure}[tb]
    \centering
    \includegraphics[width=\columnwidth]{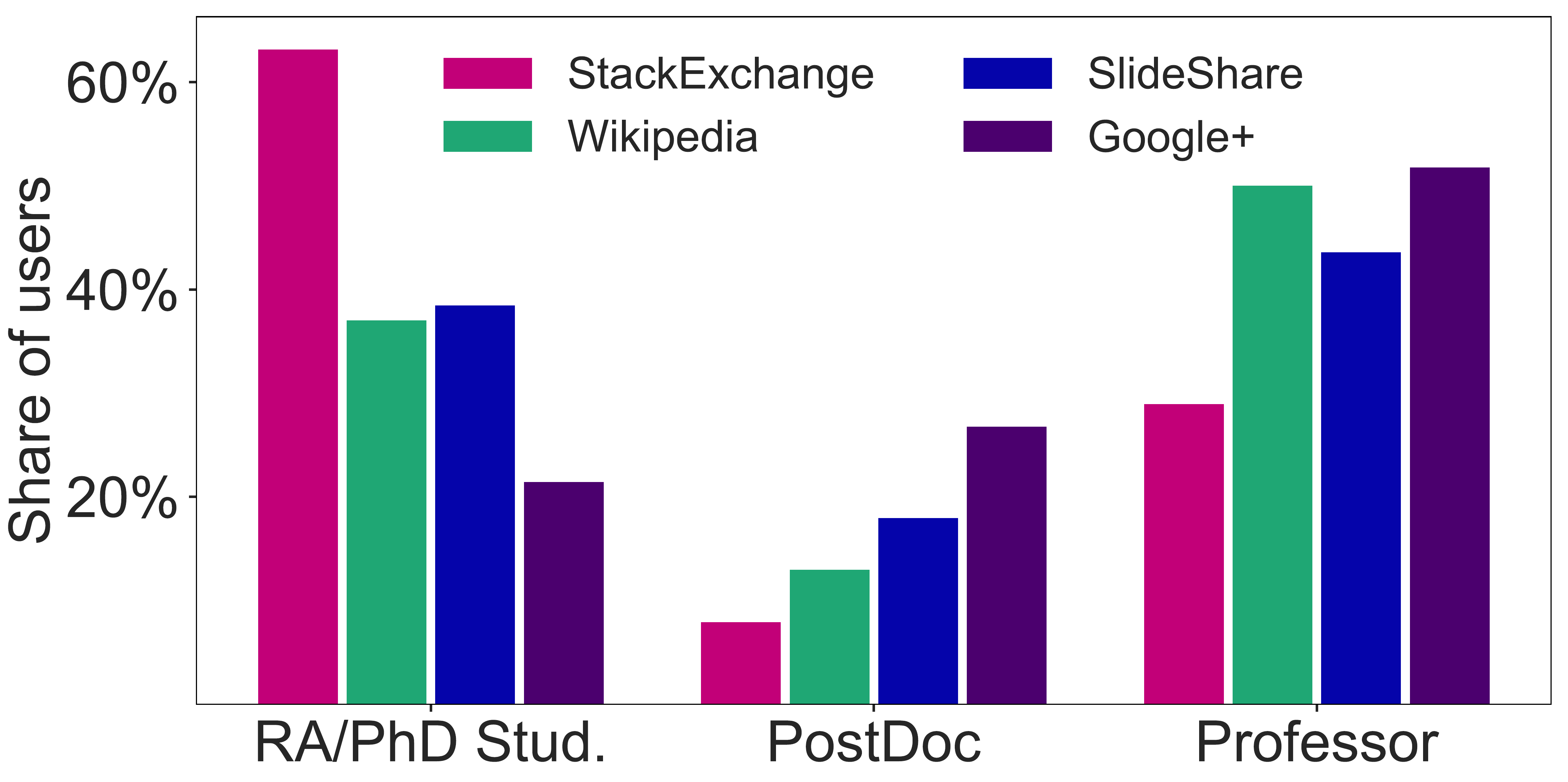}
    \caption{Role distribution of active scholarly users for selected services. \textnormal{We found strong differences between services: StackExchange is mostly used by research assistants and PhD students, while in our survey Google+, SlideShare and Wikipedia are mainly used by Professors. While the share of professors is roughly the same for Wikipedia and Google+, the share of post docs is twice-as-high for Google+, indicating a relationship between role and service use.%, showing that the demographics of active users are strongly varying across services.
    }}
    \label{fig:role_per_service}
\end{figure}

\subsection{Research experience}
To check for demographic differences between the active users of services, we plot the distribution of research experience among active users for the twelve most-frequently named services, shown in Figure~\ref{fig:exp_per_service}.
We find that services for software development and question answering -- StackExchange, StackOverflow and GitHub -- have the highest share of young researchers. On the other hand, services for networking like Google+, Twitter and LinkedIn as well as services for spreading research and information to the general public, like SlideShare and Wikipedia, have a far higher share of experienced researchers.
We identified multiple pairs of services with significant differences and large effect sizes ($>0.25$):
%The difference between Google+ and StackExchange is significant (p-value $0.00005$), as well as the difference to LinkedIn (p-value 0.0004). Google+ additionally is significantly different from StackOverflow and GitHub (p-values 0.0004 and 0.000005), and GitHub is significantly different from Wikipedia (p-value 0.0006).
The difference between Google+ and StackExchange is significant (p-value $0.00005$, effect size 0.55), as well as the difference to LinkedIn, Wikipedia and Academia.edu. Google+ additionally is significantly different from StackOverflow, GitHub and Youtube. Other pairs with significant differences are StackOverflow--LinkedIn and
GitHub--Wikipedia.
This is the first evidence that research experience influences the active scholarly use of social media. 
Altmetrics based on social media with a focus on software development will be biased towards young researchers, metrics on services mainly used for networking will be biased towards the actions of experienced researchers.
%[$\chi^2$ p < 0.001, effect size 0.55 (large)]

%\subsection{Academic roles}
To take a closer look at this finding, we compare the distribution of academic roles between services with significant differences in user experience in Figure~\ref{fig:role_per_service}. 
Google+ and Wikipedia have the highest share of experienced users. Looking at their distribution of academic roles, we see that Wikipedia has twice as many PhD students as Google+, while the latter has about twice as many postdocs compared to Wikipedia. 

Both findings indicate that different social media services fulfill different demands and thus both role and experience distributions of their users vary.

%Experience: Younger researchers code more (StackExchange, Stackoverflow, Github) while experienced researchers are active networkers (Google+, Twitter, Linkedin), share their presentations in Slideshare and contribute to Wikipedia. 

\subsection{User motivations}
\label{motiv}
%Why? LDA for discovering latent motives
%Don't tell what you do, but WHY you are doing it!
The motivations for using social media are known to vary among scholars~\cite{van2014online, conf/websci/Jordan15}. 
%However, the motivations of active scholarly users never have been researched before.
In our survey, we asked researchers to name reasons for using social media.
%Their answers allow us to uncover latent motivations for social media use, which we extract with topic models.
We ran LDA~\cite{Blei03ldaModel} on their answers to detect these latent motivations.

\begin{table}[!b] 
\caption{\textbf{Top-5 words for topics detected in the answers on ``What are other common reasons for you to like/retweet/share/... academic research on [...] services?''.} \normalfont The topics are interpretable and expose latent motives of researchers active on social media.}
\label{tab:topics}
\centering
\begin{tabu}{@{}l|l|l|l|l@{}}

\textbf{Topic 0}&
\textbf{Topic 1}&\textbf{Topic 2}&\textbf{Topic 3}&\textbf{Topic 4}
\\ 
share&relevant&find&get&interest\\
peers&others&interesting&information&interesting\\
access&research&work&new&results\\
read&think&share&topic&spread\\
people&work&knowledge&findings&content\\
\rowfont{\scriptsize}
\multicolumn{5}{l}{}\\
\rowfont{\normalfont}
\textbf{Topic 5}&
\textbf{Topic 6}&
\textbf{Topic 7}&
\textbf{Topic 8}&
\textbf{Topic 9}\\
interesting&articles&research&research&make\\
topics&download&work&share&researchers\\
show&public&relevance&like&important\\
article&news&good&academic&work\\
find&available&friends&retweet&colleagues\\
\end{tabu}

\end{table}

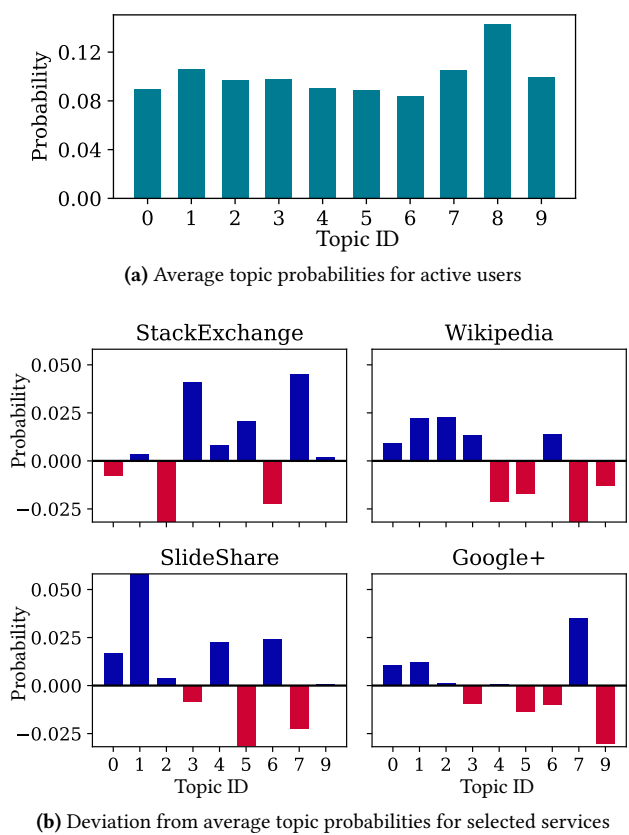
\begin{figure}[t]
\begin{subfigure}[b]{\columnwidth}
    \centering
    \begin{picture}(230,90)
    \put(0,2){\includegraphics[width=0.9\columnwidth]{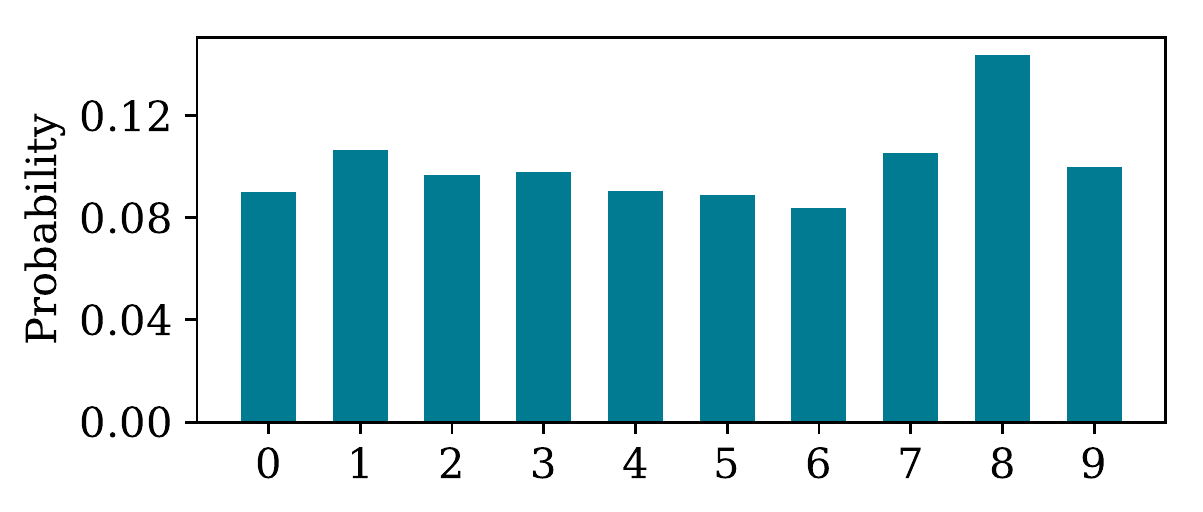}}
\put(110,0){
\normalsize{Topic ID}
}
\end{picture}
    
        \caption{\textnormal{Average topic probabilities for active users}}

    \label{fig:topics_global}
\end{subfigure}

\begin{subfigure}[b]{\columnwidth}
    \centering
    \begin{picture}(300,190)
\put(0,3){\includegraphics[width=\columnwidth]{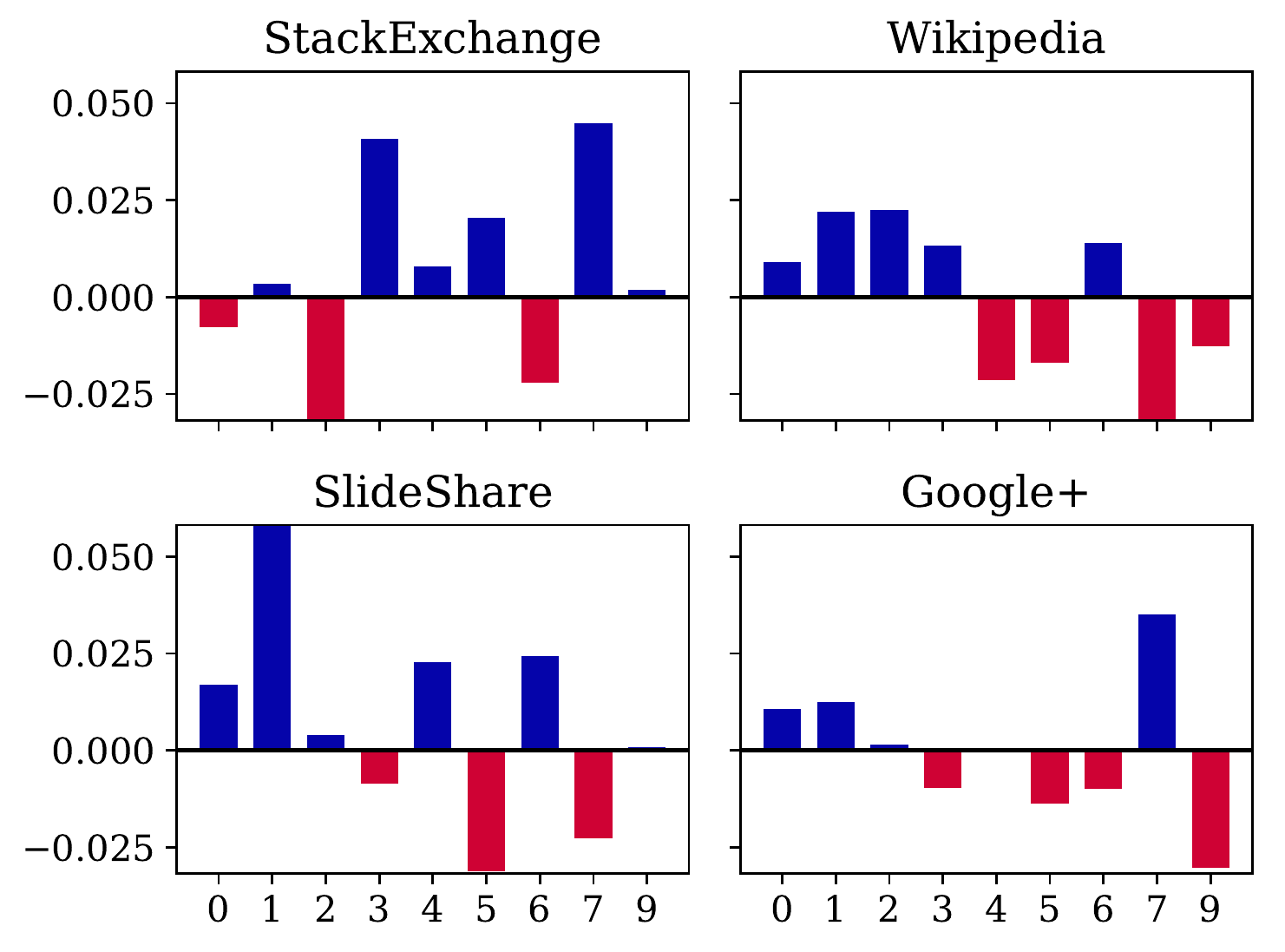}}
\put(170,0){
\small{Topic ID}
}
\put(62,0){
\small{Topic ID}
}
\put(0,35){
\begin{sideways}
\small{Probability}
\end{sideways}
}
\put(0,123){
\begin{sideways}
\small{Probability}
\end{sideways}
}
\end{picture}
    \caption{\textnormal{Deviation from average topic probabilities for selected services}}
    \label{fig:topics_per_service}
    \end{subfigure}
    \caption{Analysis of topics found in user responses on the question on reasons for using social media. \textnormal{These topics can be interpreted as latent motives and they vary for different services. The differences in motives could explain observed variations in research experience and academic roles.}}
        \label{fig:topics}
\end{figure}

Table~\ref{tab:topics} shows the detected topics. By looking at the top words of the topics and at answers with high topic probabilities in the corpus, we found that the topics can be interpreted as follows:
\textit{Topic~0}:~sharing and accessing papers of peers/other people,
\textit{Topic~1}:~users who think that their research is relevant to others,
\textit{Topic~2}:~finding and sharing interesting works,
\textit{Topic~3}:~getting information on new topics,
\textit{Topic~4}:~spreading interesting results,
\textit{Topic~5}:~showing interesting topics to the community, %finding interesting results,
\textit{Topic~6}:~downloading articles,
\textit{Topic~7}:~sharing relevant research with friends,
\textit{Topic~9}:~promoting important work of colleagues.
\textit{Topic~8}~repeats the words from the question, indicating an influence of the question on the answers. We therefore ignore this topic in our analysis.

In order to check whether there are differences between user motivations between the different services, we compare the topic distributions of the active users for different services. A user can be active in multiple services. The global topic distribution is shown in Figure~\ref{fig:topics_global}.
To find services with strong differences, we show the difference from the global topic distribution for services with a significant difference in user experience in Figure~\ref{fig:topics_per_service}.

We see that different social media services meet different needs: StackExchange has less users who want to find interesting academic works (Topic~2) and more active users who want to share research with friends (Topic~7) and to get new information (Topic~3). 
In contrast, Wikipedia has a below-average share of users who want to share relevant research with friends or their community (Topic~5 and~7), but they like to share relevant research and interesting findings with a general audience (Topic~1 and 2).
Similarly, SlideShare has an above-average share of users who use social media because they think that their research is relevant for others (Topic~1) and they like to spread interesting results (Topic~4) but have a lower probability for sharing content in their community (Topic~5 and~7).
Finally, Google+ has a higher share of users who want to share relevant research with friends but a lower probability for promoting work of colleagues (Topic~9).

These findings contribute to the understanding of the patterns found in Figure~\ref{fig:exp_per_service}: 
Google+ attracts relatively more scholars who want to share their research -- and this could explain why we see a higher share of professors~/ experienced users. 
StackExchange attracts more users who search for information -- and we can assume that this causes a high share of research assistants and PhD students, who have more practical duties and a higher need for question answering services. 

%\subsection{Semantics of user actions}
%what do we want to research?
%what do we do to research it?
%what is shown in Figure~\ref{...}?
%what is our interpretation of the patterns visible in the figure?

%In order to check for variation in the perceived semantics of activities on social media we ask the participants:
%\textit{``when you are performing the following actions, in how many cases does that indicate a positive stance on the respective target?''} with answer options \textit{always / in most cases / in few cases / never}.
%Figure~\ref{fig:positivity_per_service} shows the answers for actions from the previously analysed services, that showed significant differences in the expertise of their user base.

%Looking at the mode of each distribution, we can see that actions on different services are interpreted differently. As expected, downvoting is perceived as interacting with Wikipedia and StackExchange is perceived relatively unambiguous, while activities on SlideShare and Google+ show a more neutral attitude. The neutral attitude on SlideShare might be initiated from the fact that lower percentages of scientists can be identified on Slideshare when compared in with LinkedIn and Mendeley \cite{birkholz2013we}. Google+ is also a networking site for communicating with friend according t our finding in section \ref{motiv}.

\section{Discussion and Conclusion}
%the findings have implications...
In order to assess differences in demographics and usage motives between social media services, we studied survey responses of 1,731 active scholarly social media users. Our first analysis shows that 
\textbf{(i)~the distribution of research experience and professional roles per social service varies greatly for active users}: Experienced users use social networks and services which make research results available to the general public, young researchers are more dominant in question answering services and platforms for publishing code; and
\textbf{(ii)~the motivation of researchers for using social media services varies per service}: While services with a higher share of inexperienced researchers may attract users who search for information, services with a high share of professors~/ experienced researchers attract users who want to share their research results with friends or the general public.

%These findings have implications for the development of altmetrics and for future paper search engines:%studies which draw conclusions from scholarly social media use: 
%If we value research experience in altmetrics, actions from Google+, Twitter, Linkedin, Slideshare and Wikipedia must have a higher weight.
%Since the qualification, experience and motivations of users differs per service, the importance of measured activities is service-specific.
%The practice of uniform or simplistic weightings of measured activities in altmetrics causes bias: Actions of experienced users will not be equally influential, and users with different motivations for using %social media will be over- or under-represented.

%Additionally, the observed variety of experience and of motivations for social media usage is likely to influence the meaning of actions per service: 
%While a post mentioning a paper on StackExchange is likely a question of a young researcher (satisfying a need for information), a post mentioning a paper on Google+ is more likely explained by an experienced %researcher sharing relevant work with friends. 
%This variety should be accounted for when measuring the activities of scholars in social media for scientific impact prediction or for the ranking of papers in search engines for scholarly literature.

These findings have implications for the future development of altmetrics for scientific impact prediction: %studies which draw conclusions from scholarly social media use: 
The observed variety of experience and of motivations for social media use is likely to influence the meaning of actions per service.
While a post mentioning a paper on StackExchange is likely a question of a young researcher (satisfying a need for information), a post mentioning a paper on Google+ is more likely explained by an experienced researcher sharing a relevant publication with friends. 
This variety should be accounted for when measuring the activities of scholars in social media for scientific impact prediction, e.g.\ for improving literature search engines or the evaluation of research.% -- which benefits the scientific process.

In future work, we will extract author information from social media services and conduct surveys to better approximate the distribution of research experience per service. Combined with the distribution of citation counts depending on the research experience, this allows us to create and evaluate novel altmetrics for scientific impact prediction.
%uncover further factors that influence the quality of (alt)metrics, such as quantifying and predicting the contribution of co-authors -- an important factor for author-level metrics.

%Additionally, since the qualification, experience and motivation of users differs per service, the importance of measured activities is service-specific.
%The practice of uniform or simplistic weightings of measured activities in altmetrics causes bias: Actions of experienced users will not be equally influential, and users with different motivations for using social media will be over- or under-represented.

%Our analysis shows that current non-traditional research metrics for measuring scientific impact are biased. 
%Mechanisms to re-weight actions in altmetrics based on research experience, roles and motivations of active users must be developed, if this bias is to be prevented.

%\begin{figure}
%    \centering
%    \includegraphics[width=\columnwidth]{plots/positivity/positivity1.pdf}
%    \caption{Positivity}
%    \label{fig:positivity_per_service}
%\end{figure}

\section{Acknowledgement}
This work is part of the DFG-funded *metrics project (project number: 314727790). Further information on the project can be found on \href{http://metrics-project.net}{metrics-project.net}.

\bibliographystyle{ACM-Reference-Format}
\bibliography{acmart}

\end{document}